\documentclass[12pt]{article}
\usepackage{graphicx}
\input epsf
\usepackage{graphicx}
\usepackage{psfrag}
\usepackage{subfigure}

\usepackage{amsmath,amsfonts,amssymb,latexsym,amstext,amsbsy,amsopn,amsthm}
\begin{document}

\title{Modelling the Recoherence of Mesoscopic Superpositions in Dissipative Environments}

\author{S. G. Mokarzel$^{(1)}$, A. N. Salgueiro$^{(2*)}$ and M.C. Nemes$^{(2,3*)}$}

\maketitle

\begin{center}

{$^{(1)}$ Departamento de F\'{\i}sica--Matem\'atica,
 Instituto de F\'\i sica,
 Universidade de S\~ao Paulo, \\ C.P. 66318, 
 CEP 05315-970 S\~ao Paulo, S.P., Brazil}

{$^{(2)}$ Max Planck Institut f\"ur Kernphysik,
          Saupfercheckweg 1
          69117 Heidelberg, Germany}

{$^{(3)}$ Departamento de F\'\i sica, ICEX, 
 Universidade Federal de Minas Gerais,\\ C.P. 702, CEP 
 30161-970 Belo Horizonte, M.G. Brazil}

\end{center}

\begin{abstract}
A model is presented to describe the recently proposed experiment (J. Raimond, M. Brune and S. Haroche Phys. Rev. Lett {\bf 79}, 1964 (1997)) where a mesoscopic superposition of radiation states is prepared in a high-Q cavity which is coupled to a similar resonator. The dynamical coherence loss of such state in the absence of dissipation is reversible and can in principle be observed. We show how this picture is modified due to the presence of the environmental couplings. Analytical expressions for the experimental conditional probabilities and the linear entropy are given. We conclude that the phenomenon can still be observed provided the ratio between the damping constant and the inter-cavities coupling does not exceed about a few percent. This observation is favored for superpositions of states with large overlap.
\end{abstract}


\vfill \noindent $^*$e-mail: Andrea.Nemes.Salgueiro@mpi-hd.mpg.de \\ 
                 $^*$e-mail: carolina@fisica.ufmg.br.

\newpage
\indent
Controlling coherence properties of quantum systems has become an increasingly important subject, given the central role they play in modern technology \cite{comp} as well as in fundamental aspects of quantum theory, such as the quantum-classical transition \cite{quant}. Recently the impressive development of very refined experimental techniques opened up the possibility of testing all sorts of theoretical ideas and exploring quantum phenomena at a mesoscopic level. The constructing and monitoring of a superposition of radiation states was recently achieved the context of cavity QED \cite{exp1}. Shortly afterwards it was noted that a slight modification of that experimental setup could be used to learn about a recoherence mechanism of the same superposition of states: when the high-Q cavity containing the superposition state is coupled to another resonator, the mesoscopic quantum coherence should, in principle, first decay rapidly, then exibit sharp revivals with the period of energy exchange between the two cavities.

\indent
This idea, presented in ref. \cite{exp2}, is centered around a unitary process which introduces a new time scale related to the ``tunneling'' of the superposition between the two cavities. The well know deleterious environmental effects are completely left out of the proposal. It is the purpose of the present contribution to explicitate these effects and to give quantitative limits for the observation of the phenomenon according to the following model.

\begin{eqnarray}
\label{1}
H&=&\hbar\omega a_{1}^{\dagger}a_{1}+\hbar\omega a_{2}^{\dagger}a_{2}+\hbar \frac{\gamma}{2}\left(a_{1}^{\dagger}a_{2}+a_{2}^{\dagger}a_{1}\right)+\sum_{k}\hbar\omega_{k}b_{1k}^{\dagger}b_{1k}\nonumber \\ &&+\sum_{k}\hbar\nu_{k}b_{2k}^{\dagger}b_{2k}+\sum_{k}\left(\hbar\beta_{1k}b_{1k}^{\dagger}a_{1}+h.c.\right)+\sum_{k}\left(\hbar\beta_{2k}b_{2k}^{\dagger}a_{2}+h.c.\right)
\end{eqnarray}

\noindent 
The first two terms on the r.h.s. of the eq.(\ref{1}) stand for the two resonators. Their coupling is given by the third term of the r.h.s. of the same equation. This choice for the coupling is based on the fact that the time evolution of a superposition of coherent states remains a superposition of coherent states at later times under the dynamics given by the first three terms on the r.h.s. of eq.(\ref{1}) \cite{omnes}. The inclusion of nonresonant terms in this coupling would therefore completely destroy the simple picture proposed in ref.(\cite{exp2}). The presence of an environment and its coupling to the two resonators is modelled by the standard collection of harmonic oscillators (with frequencies $\omega_{k}(\nu_{k})$ in the cavity $1(2)$) interacting separately with the two resonators. The couplings are again of the Rotating Wave Approximation (RWA) form which is well justified in this context \cite{RWA}. In eq.(\ref{1}) $\alpha_{1k}$ and $\alpha_{2k}$  stand for the coupling constants.

\indent
The dynamics of the full system described by eq.(\ref{1}) obeys Schr\"odinger's equation and the corresponding state vector is a pure state $|\psi(t)\rangle$. Since we are interested in the dynamics of systems $1$ and $2$ only, we deduce a master equation from eq.(\ref{1}) by means of the usual Born-Markov approximation and get for $\rho_{s}(t)=Tr_{env}\left(|\psi(t)\rangle\langle\psi(t)|\right)$,

\begin{eqnarray}
\label{2}
i\frac{d\rho_{s}(t)}{dt}&=&\left(-i\omega-k\right)a_{1}^{\dagger}a_{1}\rho_{s}+
\left(-i\omega-k\right)\rho_{s}a_{1}^{\dagger}a_{1}+2ka_{1}\rho_{s}a_{1}^{\dagger}\nonumber \\&&+\left(-i\omega-k\right)a_{2}^{\dagger}a_{2}\rho_{s}+
\left(-i\omega-k\right)\rho_{s}a_{2}^{\dagger}a_{2}+2ka_{2}\rho_{s}a_{2}^{\dagger}\nonumber \\&&-i\gamma\left(a_{1}^{\dagger}a_{2}\rho_{s}-\rho_{s}a_{1}^{\dagger}a_{2}\right)-i\gamma\left(a_{1}a_{2}^{\dagger}\rho_{s}-\rho_{s}a_{1}a_{2}^{\dagger}\right)
\end{eqnarray}

\noindent
where we have defined $k\equiv D_{1}(\omega){|\alpha_{1}(\omega)|}^{2}=D_{2}(\omega){|\alpha_{2}(\omega)|}^{2}$ to be the damping constants of the cavities; $D_{1,2}(\omega)$ stands for the density of states at the resonator's frequency $\omega$, and the continumm limit has been taken with respect to the environmental frequencies.

\indent
Note that the reduced density which obeys eq.(\ref{2}) describes the two cavities. The solution of the analogous problem involving only one cavity can be found in several text books. The novel feature here are the terms involving operators of both cavities (e.g. the last terms on the r.h.s. of eq.(\ref{2})) whose physical origin is the coupling between the two resonators. It is however also possible in this case to find an analytical solution by noting that the set of all superoperators in the above equation form a Lie-algebra (\cite{sonia}). For the initial condition of interest (see ref.\cite{exp1}),

\begin{equation}
\label{3}
\rho_{s}(0,{g \atop e})=\frac{1}{N_{{g \atop e}}}\left(e^{-i\phi}{|\alpha(0)e^{-i\phi}\rangle}_{1}\pm|\alpha(0)e^{i\phi}{\rangle}_{1}\right)(h.c.)\otimes{|0\rangle}_{22}{\langle}0|
\end{equation} 

\noindent
we get

\begin{eqnarray}
\label{4}
\rho_{s}(t,{g \atop e})&=&\frac{1}{N_{{g \atop e}}}\Bigl\{|\alpha_{1}^{(+)}(t)\rangle\langle \alpha_{1}^{(+)}(t)| \otimes |\alpha_{2}^{(+)}(t)\rangle\langle \alpha_{2}^{(+)}(t)|+ |\alpha_{1}^{(-)}(t)\rangle\langle \alpha_{1}^{(-)}(t)| \otimes |\alpha_{2}^{(-)}(t)\rangle\langle\alpha_{2}^{(-)}(t)|\nonumber \\  
&&\pm\bigl\{e^{\bigl[-i\phi+\frac{1}{2}\left({|\alpha_{1}^{(+)}(t)|}^{2}+{|\alpha_{1}^{(-)}(t)|}^{2}+{|\alpha_{2}^{(+)}(t)|}^{2}+{|\alpha_{2}^{(-)}(t)|}^{2}-2{|\alpha(0)|}^{2}\right)\bigr]}\nonumber \\  &&e^{\bigl[\alpha_{1}^{(+)}(0){{\alpha}^{*}_{1}}^{(-)}(0)- \alpha_{1}^{(+)}(t){{\alpha}^{*}_{1}}^{(-)}(t)-\alpha_{2}^{(+)}(t){{\alpha}^{*}_{2}}^{(-)}(t)\bigr]}\nonumber \\ &&|\alpha_{1}^{(+)}(t)\rangle\langle\alpha_{1}^{(-)}(t)|\otimes|\alpha_{2}^{(+)}(t)\rangle\langle\alpha_{2}^{(-)}(t)|+h.c.\bigr\}\Bigr\}
\end{eqnarray}

\noindent
where the letters $g$ and $e$ are related to the two signs $\pm$ in the above equations. They correspond, in the experiment of ref.\cite{exp1}, to measuring the first atom in the state $g$ (or $e$) and leaving in the high-Q cavity $C_{1}$ an odd or even ``cat state''. Also in order to obtain eq.(\ref{4}), we assume $\gamma>>k$ (this is not necessary to obtain the analytical solution but corresponds to the physical situation in case) and get

\begin{eqnarray}
\label{5}
\alpha_{1}^{(\pm)}(t)&=&\alpha(0)e^{\pm i\phi}e^{-(k+i\omega)t}\cos{(\frac{\gamma t}{2})}
\nonumber \\ \alpha_{2}^{(\pm)}(t)&=&2i\gamma\alpha(0)e^{\pm i\phi}e^{-(k+i\omega)t}\sin{(\frac{\gamma t}{2})}
\end{eqnarray}

\noindent
where $\alpha(0)$ is the initial amplitude of the coherent field in cavity $1$ and 
\[
N_{{g \atop e}}=2[1\pm e^{-{|\alpha(0)|}^{2}\left(1-\cos(2\phi)\right)\cos(\phi+{|\alpha(0)|}^{2}\sin(2\phi))}].
\]

\indent
In order to describe the dynamics of cavity $1$ alone, one can trace out the degrees of freedom associated with the index $2$

\begin{eqnarray}
\label{7}
\rho_{1}(t,{g \atop e})&=&Tr_{2}(\rho_{s}(t,{g \atop e})) \nonumber \\
&&= \frac{1}{N_{{g \atop e}}}\Bigl\{ |\alpha_{1}^{(+)}(t)\rangle\langle\alpha_{1}^{(+)}(t)|+ |\alpha_{1}^{(-)}(t)\rangle\langle\alpha_{1}^{(-)}(t)|\nonumber \\ &&\pm \bigl\{
e^{\bigl[-\frac{1}{2}D^{2}(\alpha(0)e^{-i\phi},\alpha(0)e^{i\phi})\left(1-e^{-2kt}{\cos}^{2}(\frac{gt}{2})\right)\bigr]}e^{\bigl[-i\phi+{|\alpha(0)|}^{2}(1-e^{-2kt}{\cos}^{2}(\frac{gt}{2}))\sin(2\phi)\bigr]}\nonumber \\ &&|\alpha_{1}^{(-)}(t)\rangle\langle\alpha_{1}^{(+)}(t)|+h.c.\bigr\}\Bigr\} 
\end{eqnarray}

\noindent
where $D(\alpha(0)e^{-i\phi},\alpha(0)e^{i\phi})$ is the distance between the states in the superposition and is given by $D(\alpha,\beta)=|\alpha-\beta|$. Note here that if we take the limit $\gamma=0$ we recover the usual reduced density matrix of a superposition in a dissipative environment \cite{luis,nos}. Turning on the coupling between the two cavities brings in a new time scale in the problem, i.e., the characteristic time for energy exchange between the two cavities. The interesting point is that this latter time dependence is periodic and therefore completely different from the exponential which characterizes decoherence in general. Of course if these two time scales are sufficiently different the proposed ``reversibility'' of decoherence might be observed. We now turn to the quantitative question of how different $\gamma$ and $k$ have to be in order for the phenomenon to be observable. For this purpose we calculate the conditional probability $P_{ee}(t)$ and $P_{ge}(t)$ to detect the second atom in state $e$ provided the first was detected in $e(g)$. Using eq.(\ref{7}) these probabilities are given by

\begin{equation}
P_{{ee \atop ge}}(t)=\frac{1}{2}\{1-Re[e^{-i\phi}Tr_{F}[e^{-2i\phi a^{\dagger}a}\rho_{1}(t,{g \atop e})]]\}
\end{equation}

\noindent
or, explicitely, 

\begin{eqnarray}
\label{8}
P_{{ee \atop ge}}(t)&=&\frac{1}{2}-\frac{1}{2N_{{g \atop e}}}\biggl\{2e^{-D^{2}(\alpha_{1}^{(-)}(t),\alpha_{1}^{(+)}(t))}\cos\bigl[\phi+{|\alpha(0)|}^{2}e^{-2kt}{\cos}^{2}(\frac{\gamma t}{2})\sin(2\phi)\bigr]\nonumber \\ &&\pm e^{-\frac{1}{2}D^{2}(\alpha(0)e^{-i\phi},\alpha(0)e^{i\phi})(1-e^{-2kt}\cos^{2}(\frac{\gamma t}{2}))}\Bigl\{\cos\bigl[{|\alpha(0)|}^{2}(1-e^{-2kt}\cos^{2}(\frac{\gamma t}{2}))\sin(2\phi)\bigr]\nonumber \\ &&+e^{-\frac{1}{2}D^{2}(\alpha_{1}^{(-)}(t)e^{-i\phi},\alpha^{(+)}_{1}(t)e^{i\phi})}\cos\bigl[2\phi+{|\alpha(0)|}^{2}\sin(2\phi)\bigl(1-e^{-2kt}\cos^{2}(\frac{\gamma t}{2})+2e^{-2kt}\cos^{2}(\frac{\gamma t}{2})\bigr)\bigr]\Bigl\}\biggr\} \nonumber \\
\end{eqnarray}

The correlation signal $\eta$ is defined as $\eta(t)=P_{ee}(t)-P_{ge}(t)$. Also in order to characterize decoherence we calculate the linear entropy \cite{zurek,griffin} $\delta(t)=1-Tr(\rho_{1}^{2}(t,{g \atop e}))$.

\begin{eqnarray}
\label{9}
\delta(t)&=&\frac{2}{N^{2}_{{g \atop e}}}\Bigl\{{(1+e^{(-\frac{1}{2}D(\alpha(0)e^{-i\phi},\alpha(0)e^{i\phi}))})}^{2} \nonumber \\ 
&&-{\bigl(e^{(-\frac{1}{2}D(\alpha_{1}^{(-)}(t),\alpha_{1}^{(+)}(t)))}+e^{(\frac{1}{2}D(\alpha_{1}^{(-)}(t),\alpha_{1}^{(+)}(t))-\frac{1}{2}D(\alpha(0)e^{-i\phi},\alpha(0)e^{i\phi}))}\bigr)^{2}}\Bigr\}
\end{eqnarray}

\noindent
where $\alpha_{1}^{(\pm)}$ are defined in eq.(\ref{5}).
Since the expression for the correlation signal is rather lengthy and not very illuminating we simplify it in the small overlap limit, i.e., $\langle\alpha(0)e^{-i\phi}|\alpha(0)e^{i\phi}\rangle<<1$. In this case we get for the correlation signal and the following simple expression

\begin{equation}
\label{10}
\eta(t)\approx\frac{1}{2}\cos\Bigl[|\alpha(0)|^2\bigl(1-e^{-2kt}{\cos}^{2}(\frac{\gamma t}{2})\bigr)\sin(2\phi)\Bigr]e^{\bigl[-\frac{1}{2}D^{2}(\alpha(0) e^{-i\phi},\alpha(0) e^{i\phi})(1-e^{-2kt}{\cos}^{2}(\frac{\gamma t}{2}))\bigr]}
\end{equation}

\noindent
Note that both expressions contain the ``distance'' factor which is directly related to decoherence, as can be seen in eqs.(\ref{9}) and (\ref{10}). The ${\cos}^{2}(\frac{\gamma t}{2})$ factor contained in $\alpha^{(\pm)}_{1}$ in the expresssion for the linear entropy is responsible for the recoherence phenomenon. The same factor is present in the correlation signal, which in addition has a term like $\cos[|\alpha(0)|^2(1-e^{-2kt}{\cos}^{2}(\frac{\gamma t}{2}))\sin(2\phi)]$. This term, provided $k<<\gamma$ will oscillate with a period corresponding to the energy exchange period between the two cavities.

\indent 
In figures $1$ and $2$ we show the correlation signal and linear entropy for the case when the coherent states in the superposition are distant, in the sense that the small overlap approximation is valid ($\phi=0.98rad$). In figure $1$ we clearly see the recoherence peaks in the case of no dissipation, as in ref.\cite{exp2}. The effect of dissipation, which can also be appreciated in this figure is, roughly speaking, to superimpose a decaying exponential on this curve. For the value in the figure $k/\gamma=0.01$ the second peak
is still about $70\%$ in intensity with respect to the first one. Already for $k/\gamma=0.05$ it has disappeared. The revival observed in this figure are also revealed in the linear entropy. It starts at zero indicating a pure state and attains its maximum when the correlation signal is a minimum. At this time the state in cavity $C_{1}$ corresponds to a statistical mixture. The fact that the maximum value of $\delta(t)$ is $0.5$ in this case is related to the small overlap in the component of the initial superposition. Given the form of the coupling, one sees from eq.(\ref{4}) that the state generated in the second cavity $C_{2}$ will be very similar to the one in the first and therefore we will also have a small overlap. In this case the reduced density $\rho_{1}$ will approximately be a statistical mixture. This is clearly different in the cases of figures $3$ and $4$ where we have a larger overlap $\phi=0.4rad$ between the two components of the superposition. Other than that the figures are similar with similar interpretations. The significant difference resides in the fact that the effect of decoherence is slower in this case. In figure $5$ we show that the second peak in the correlation signal can still be seen with reasonable intensity for $k/\gamma=0.05$.

\noindent
From the model we presented here we conclude that the ``reversible decoherence'' proposed in ref.\cite{exp2} can in fact be observed provided: a-) the coupling between the two cavities is well modelled by $\gamma(a^{\dagger}_{1}a_{2}+ a_{1}a^{\dagger}_{2})$, as discussed in the text. b-) The ratio $k/\gamma$ should not exceed a few percent and the distance between the components of the superpositions will definitely influence this value. 

\vspace{.3cm}

\noindent {\bf Acknowledgements.} We acknowledge fruitful discussions with profs. H.A. Weidenm\"uller and S. Haroche. This work is supported by Conselho Nacional de Desenvolimento Cient\'{\i}fico e Tecnol\'ogico (CNPq) (SGM and MCN), Funda\c{c}\~ao de Amparo \`a Pesquisa do Estado de S\~ao Paulo (FAPESP)(ANS) and the Humboldt Foundation (MCN).


\newpage

\section*{Figure captions.}

\noindent Fig. 1 - Correlation signal $\eta(t)$ for $k/\gamma=0$ (full line) and $k/\gamma=0.01$ (dashed line) for ${|\alpha(0)|}^{2}=3.3$ and $\phi=0.98rad$. 

\vspace{.5cm}

\noindent Fig. 2 - Same as Figure 1 for the linear entropy $\delta(t)$.

\vspace{.5cm} 

\noindent Fig. 3 -  Correlation signal $\eta(t)$ for $k/\gamma=0$ (full line) and $k/\gamma=0.01$ (dashed line) for ${|\alpha(0)|}^{2}=3.3$ and $\phi=0.4rad$.

\vspace{.5cm}

\noindent Fig. 4 - Same as Figure 3 for the linear entropy $\delta(t)$.

\vspace{.5cm}

\noindent Fig. 5 - Correlation signal for the case in figure 3, but $k/\gamma-0.05$.

\vspace{.5cm}

\noindent Fig. 6 - Linear entropy for the case in figure 3, but $k/\gamma-0.05$.

\newpage

\psfrag{eta(t)}[][]{$\eta(t)$}
\psfrag{delta(t)}[][]{$\delta(t)$}
\psfrag{(gt/2pi)}[][]{$(\gamma t/ {2\pi})$}

\begin{figure}[ht]
\centering
\unitlength1cm  
\begin{picture}(14,10)
\put(0.5,6.5){$\eta(t)$}
\put(10,0.5){$(\frac{gt}{2\pi})$} 
\epsfysize=6cm
\epsffile{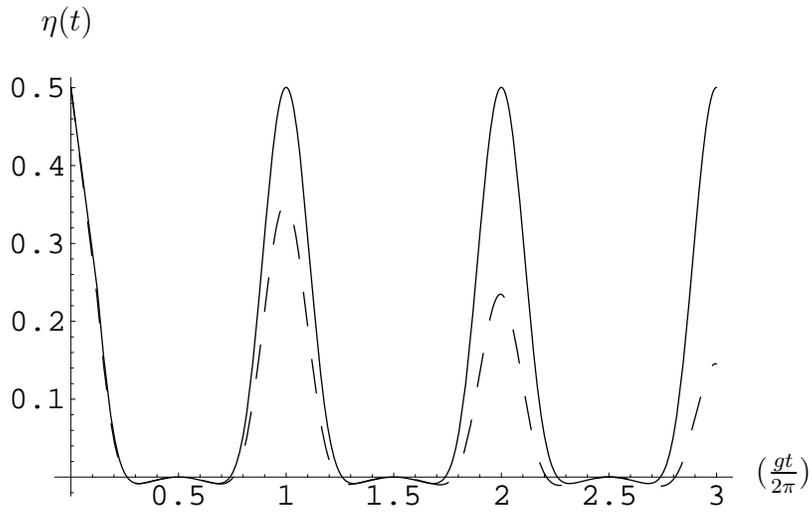} 
\end{picture}
\caption{Correlation signal $\eta(t)$ for $k/\gamma=0$ (full line) and $k/\gamma=0.01$ (dashed line) for ${|\alpha(0)|}^{2}=3.3$ and $\phi=0.98rad$.}
\label{fig1}
\end{figure}

\begin{figure}[ht]
\centering
\unitlength1cm  
\begin{picture}(14,10)
\put(0.5,6.5){$\delta(t)$}
\put(10,0.5){$(\frac{gt}{2\pi})$} 
\epsfysize=6cm
\epsffile{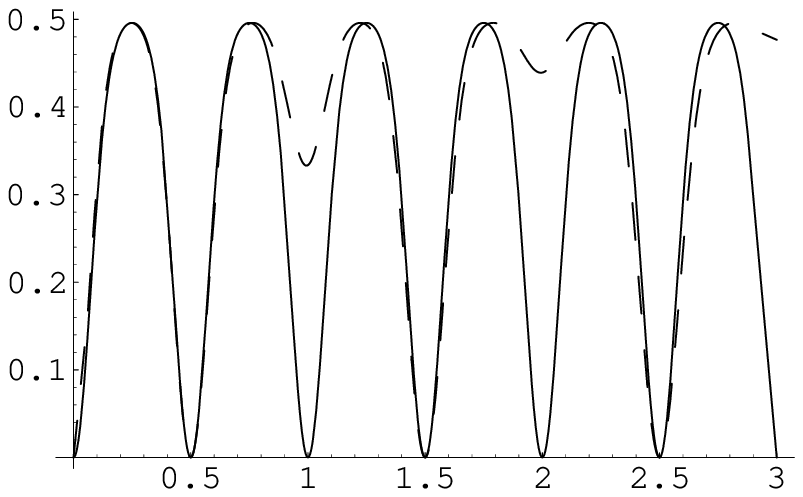} 
\end{picture}
\caption{ Same as Figure 1 for the linear entropy $\delta(t)$.}
\label{fig2}
\end{figure}

\begin{figure}[ht]
\centering
\unitlength1cm  
\begin{picture}(14,10)
\put(0.5,6.5){$\eta(t)$}
\put(10,0.5){$(\frac{gt}{2\pi})$} 
\epsfysize=6cm
\epsffile{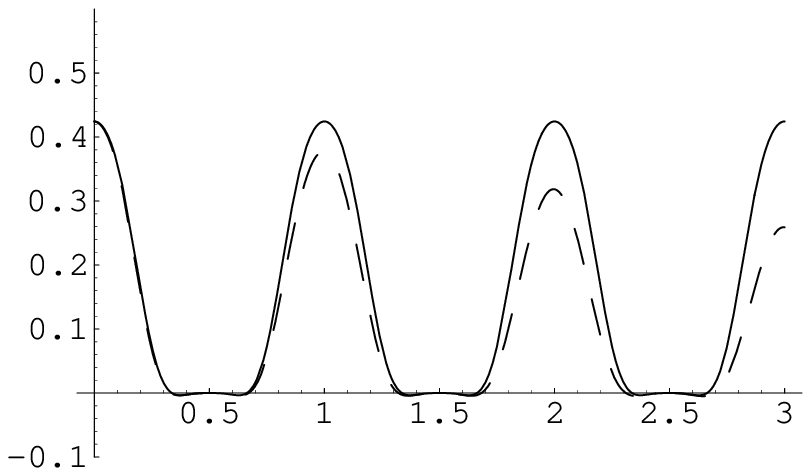} 
\end{picture}
\caption{Correlation signal $\eta(t)$ for $k/\gamma=0$ (full line) and $k/\gamma=0.01$ (dashed line) for ${|\alpha(0)|}^{2}=3.3$ and $\phi=0.4rad$.}
\label{fig4}
\end{figure}

\begin{figure}[ht]
\centering
\unitlength1cm  
\begin{picture}(14,10)
\put(0.5,6.5){$\delta(t)$}
\put(10,0.5){$(\frac{gt}{2\pi})$} 
\epsfysize=6cm
\epsffile{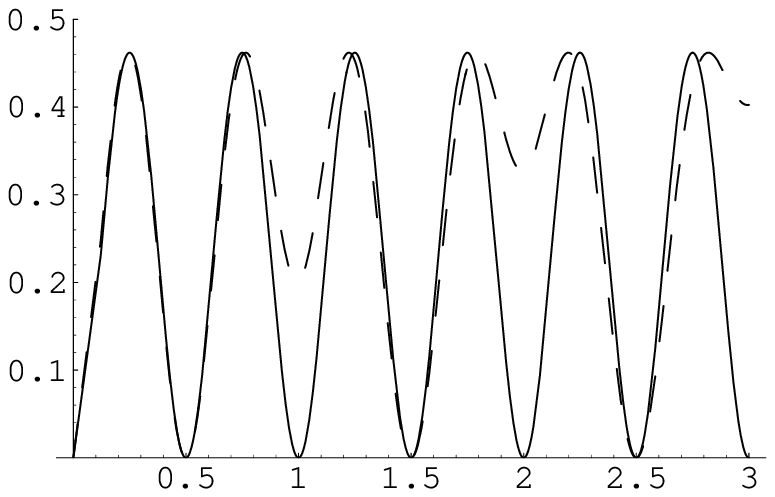} 
\end{picture}
\caption{Same as Figure 3 for the linear entropy $\delta(t)$.}
\label{fig5}
\end{figure}

\begin{figure}[ht]
\centering
\unitlength1cm  
\begin{picture}(14,10)
\put(0.5,6.5){$\eta(t)$}
\put(10,0.5){$(\frac{gt}{2\pi})$} 
\epsfysize=6cm
\epsffile{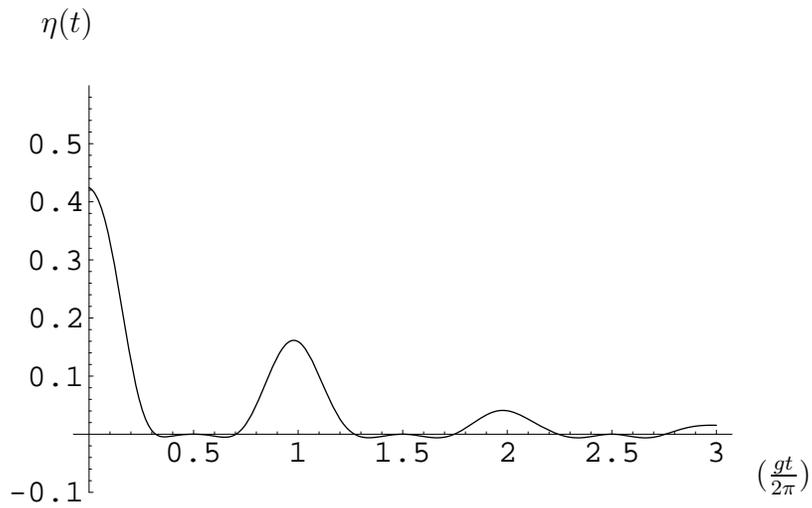} 
\end{picture}
\caption{Correlation signal for the case in figure 3, but $k/\gamma-0.05$.}
\label{fig6}
\end{figure}

\begin{figure}[ht]
\centering
\unitlength1cm  
\begin{picture}(14,10)
\put(0.5,6.5){$\delta(t)$}
\put(10,0.5){$(\frac{gt}{2\pi})$} 
\epsfysize=6cm
\epsffile{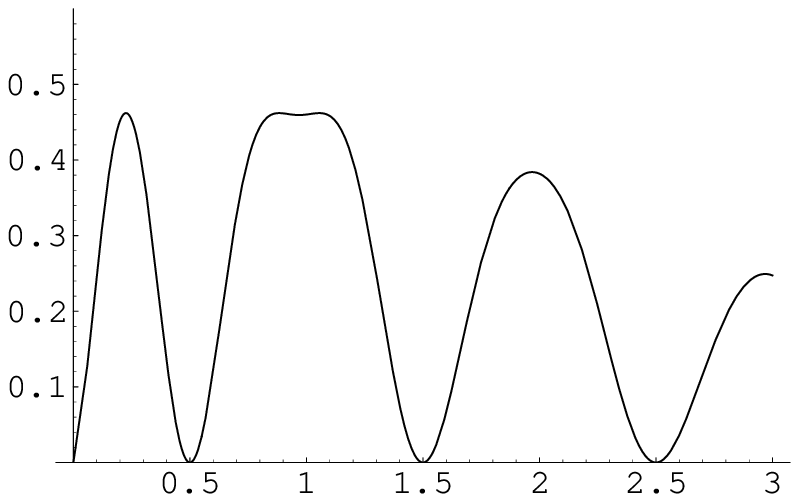} 
\end{picture}
\caption{Linear entropy for the case in figure 3, but $k/\gamma-0.05$.}
\label{fig7}
\end{figure}

\end{document}